# A Report on the Workshop on Biobanking Informatics

Jitendra Jonnagaddala
Translational Cancer Research Network
Prince of Wales Clinical School
University of New South Wales (UNSW)
Sydney, Australia

Damian Sue
Asia-Pacific Ubiquitous Healthcare Research Centre
Australian School of Business
University of New South Wales (UNSW)
Sydney, Australia

*Abstract*—The Workshop on Biobanking Informatics in NSW 2013 (WBIN13) was held on Friday, 10 May 2013 at The Wallace Wurth Building in the University of New South Wales. This report summarises the keynotes, presentations and discussions in WBIN13 which discusses current research in the field of Biobanking Informatics in Australia and internationally

*Keywords—Biobanking, Informatics, SOP, integration, workshop*

## I. Introduction

The Workshop on Biobanking Informatics in NSW 2013 (WBIN13) was organised by the Translational Cancer Research Network (TCRN) to bring together practitioners and researchers (Australian and International) with a broad range of expertise in the field of Biobanking Informatics to provide a forum for the exchange of experiences, ideas and approaches. Details about the workshop can be found at https://wbin13.eventbrite.com.au. The workshop took place on the 10th of May 2013 in the Wallace Wurth Building at the University of New South Wales with Professor Nicholas Hawkins moderating the session.

The aims of the workshop were as follows (1) To assess and compare current research in the field of Biobanking Informatics in Australia and internationally (2) To identify knowledge gaps and priority areas to be addressed (3) To explore the possibility for cross-disciplinary collaboration and areas for future joint development (4) To provide a forum to foster the development of joint research programs and student/staff exchange. Prof. Nick kicked off the workshop with discussion on biobanking and its place in translational research (Figure 1)[1].

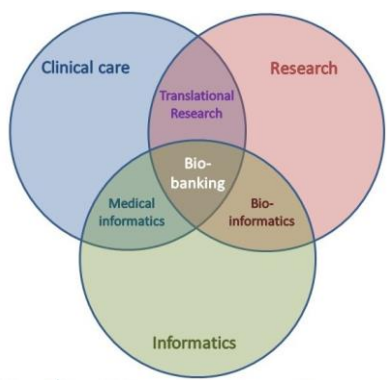

*Figure 1: Biobanking and translational research*

The workshop was separated into a morning and evening session. It comprised of ten speeches, and two panel discussions. The first keynote speech was delivered in the morning which discussed "Cancer Research Information Cloud at Taipei Medical University, Taiwan". After Morning Tea two more speeches were conducted, the first was on caTissue an open source Biobanking Laboratory Information Management System (LIMS), and the second was on integrating Biospecimen Management Systems with Anatomical Pathology Laboratory Information Systems (LIS) and Digital Pathology Systems. After lunch the Second Keynote speech was conducted on the US perspectives on biobanking which was followed by a small panel discussion. Finally, the workshop concluded with a Seminar, which comprised of short presentations from local and state perspectives on linking cancer biobanks in NSW as well as a final Panel Discussion.

## II. Morning Session

### A. First Speech (Keynote): Cancer Research Information Cloud at Taipei Medical University, Taiwan

Professor Jack Li is a pioneer of Medical Informatics research in Taiwan and is the Vice President of Taipei Medical University as well as an adjunct professor of Graduate Institute of Biomedical Informatics, and President of the Asia Pacific Association for Medical Informatics. Professor Li discussed the benefits of medical cloud storage for bio-specimen data which integrates biospecimen samples from multiple sites.

The presenter also discussed Taipei Medical University's implementation of this cloud system for their biobank sites and emphasized the benefits of providing all biospecimen data online which includes the ease of access to biospecimen samples for researchers, the improved efficiency for ethics committees to approve the use of biospecimens samples as it is done online rather than paper based methods, and the possibility of using Business Intelligence tools such as Google Motion Chart to analyze data, now that it has been collected in one location. Another major benefit to cloud based systems is unlimited storage in the cloud. It was found that local systems had no capacity to store all biospecimen data, and old records had to be deleted to "make room" for new data. This resulted in the loss of historical data which could be used for further data analysis.

The system should only hold de-identified data which is important in protecting patient confidentiality. Also, every EMR

record requires consent from ethics committee before being published on the cloud. Additionally, all sites are required to harmonize their Standard Operating Procedures (SOP) for Cloud based systems to be possible. Professor Li suggested other countries, including Australia, to adopt a similar cloud based approach to biospecimen samples (Figure 2).

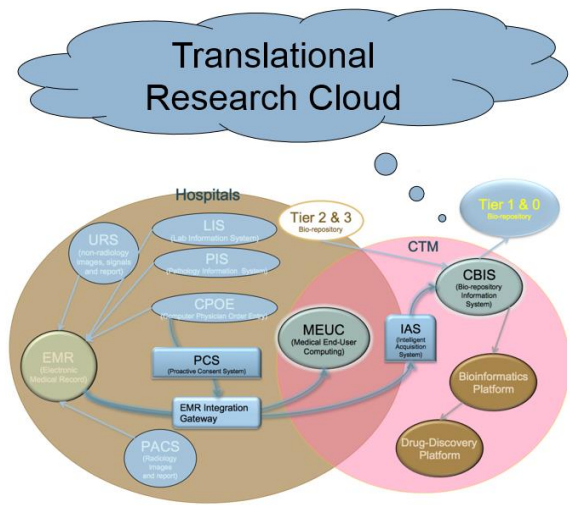

*Figure 2: Translational Research Cloud*

### B. Second speech: caTissue+

Srikanth Adiga from Krishagni Solutions Pty Ltd discussed the program caTissue+ which is an open source Biobanking Laboratory Information Management System (LIMS) and the possibilities it can provide for integrating biobanks. It has been adopted by over 50 biobanks across the globe and, being an open source system, has various information/guides available on the web.

### C. Third speech: Integrating Biospecimen Management System with Anatomical Pathology Laboratory Information System (LIS) and Digital Pathology Systems

Jitendra Jonnagaddala, an Information Manager at TCRN presented Integrating Biospecimen Management System with Anatomical Pathology Laboratory Information System (LIS) and Digital Pathology Systems. The presenter stressed the importance of integrated clinical and research systems being scalable, flexible, and interoperable. The system developed must be generic to allow more information to be loaded into the system in the future, as well as allowing additional data types to be entered into the system[2].

The pilot project between Lowy Bio-repository, SEALS, and UNSW School of Medical Sciences was also discussed, as well as the lessons learned from the integration experience. The pilot project linked Aperio (a digital pathology system) and OmniLab (Laboratory Information System) using caTissue+. Challenges found in integrating the two systems was that the standard operating procedures (SOPS) from both Lowy and SEALS were extremely different and had to be integrated. Also another hurdle was convincing people to use the system. It was also found that it was difficult to determine who to talk to when obtaining permission and data for Biospecimens. Furthermore, it was difficult in gaining trust from the hospital to obtain biospecimen data.

caTissue+ was found to be very flexible and integration between the two systems would not have been possible without it [3]. Moreover, caTissue+ provides the opportunity for other systems to be integrated with the existing system. Any system using HL7 could be integrated with caTissue+.

However, it was also mentioned that having completed this project and understanding which people/organizations need to be consulted in order to obtain the necessary authority to integrate systems. Similar projects in the future can be completed in a much shorter timeframe.

### D. Fourth Speech (Keynote): US perspective

Dr. Jim Vaught is the Deputy Director of the U.S. National Cancer Institute's (NCI) Office of Biorepository and Biospeicmen Research Banks (BBRB). Dr. Vaught discussed several issues (1) biobanking business planning and sustainability options (2) international sample/data access and research collaboration and (3) Translational and economic impacts of biobanking.

The speech began with the various issues faced in the United States with collecting bio-specimens from various sites and also in obtaining permission required to use the samples and how Australia could learn from it when integrating our systems[4]. NCI best practices for biospecimen resources and the adoption by others was considered critical in getting good quality samples as studies showed there were discrepancies between biobanks in storing samples (i.e. samples stored at different temperatures) which led to many samples being collected that were unusable[5]. It was found that once NCI best practices were implemented, sample quality did indeed improve. It is noted that it has not been established whether it is feasible for implementing best practices yet.

Another major issue found in the US model was obtaining permission for the biospecimen samples[6]. Different samples were owned by different centers, and sometimes multiple centers. To overcome this, a workshop was required to get all relevant stakeholders to collaborate and produce a standardized best practices. NSW needs to achieve something similar to unify our centers.

### E. Fifth Speech: using CTRNet approach to link biobanks though education

Dr. Peter Watson is a Professor in the Department of Pathology, BC Cancer Agency and the University of British Columbia who presented "The CTRNet (Canadian Tumour Repository Network) approach to linking biobanks through education and support"[7]. He began by describing the quality issues faced in Canada due to a lack of standardization in SOP which resulted in a lack of capacity to meet demand, and reduced efficiency. The response was to create a certification program where all biobanks certified has undertaken fundamental training and are approved by the ethics board in Canada. All biobanks were to use the same SOP, and had shared governance between pathology and biobanks. This certification process has improved the quality of biospecimens collected and the program has been developed for international usage.

## F. Sixth Speech: opportunities for Australian biobanks

Dr. Nik Zeps is a PhD scientist from the University of Western Australia who presented "opportunities for Australian biobanks". Dr. Zeps pointed out that that Western Australia also faced similar problems to Canada in that many samples collected by biobanks were not up to standard. He also discussed many issues with Australian biobanks, in that they are fragmented, inefficient and present difficulties for access and recruitment. He also discussed the Cancer Council's findings that a majority of Australian biobanks are solely tissue based, have poor linkage to medical data, have limited linkage to lifestyle data, are too small to be research effective, and have not been designed as a truly open source of access. Also there was difficulty in gaining consent and ethical approvals due to the ethics board requiring a comprehensive statement of approval, rather than a simple statement from the patients permitting the use of their tissue[8-10].

Dr. Zeps concluded his speech with suggesting Australia should adopt a plan to have our systems centralized (data collected from local sites and sent to the hub), distributed (local nodes manage collection, processing and storage until required) and networked (with independent biobanks integrated by informatics solutions)[11].

It was also stressed how a scientific approach should be taken towards Informatics rather than an opinion based approach and that the Canada model of networking is key to sharing data between biobanks.

## G. Panel Discussion

A panel discussion comprised of 3 people, Dr. Jim Vaught, Dr. Peter Watson and Dr. Nik Zeps was then conducted. Several questions were asked, including how we can deal with the legacy issue of biobanks, and governance issues. There was also a healthy discussion on methods of convincing the Australian public that biobanks are important and should be funded as well as the ideal governance structure for Australian biobanks.

## III. EVENING SESSION

After the panel discussion the Seminar on "Linking cancer biobanks in NSW and what can we learn from others" was conducted with Professor Nicholas Hawkins once again as the Moderator. The session was comprised of a series of short presentations of local and state perspectives on the benefits of interoperability between biobanks at an information level. It was then followed by an interactive panel discussion that drew on international experiences in this area.

## A. Seventh speech: The NSW State Experience

Professor Anna deFazio is the Head of the Gynaecological Oncology Research Group at the Westmead Millennium Institute and a Professor at the Sydney West Translational Cancer Research Centre. Professor deFazio presented the NSW experience of getting interoperability between biobanks and highlighted the difficulty in linking biobanks with the different translational research centres that have been established with the cancer institute of NSW especially as there are multiple types of biobanks which include multi biobanks, and biobanks with multiple nodes[12, 13].

Professor deFazio first discussed the Australian ovarian cancer study which recruited 1859 women with ovarian, fallopian tube, and primary peritoneal cancer from around Australia. All samples were collected using standardized operation procedures and project contributed to over 120 publications since 2007 and more than 90 projects approved.

From very beginning it was known that it would be costly to undertake this project. The cost to use each biospecimen array was over $1000 in early days. It was therefore essential to ensure that these specimens had high quality metadata. The question was how to collect the data from many different databases in an efficient, yet effective method. In terms of access, hospitals and clinics were willing to provide access to their biospecimen databases. A common issue was found that much of the data within the databases were incomplete due to a lack of SOP as well as data on the same patient being spread between multiple centers. It was found that some databases did not even have a data dictionary. The problem was that there was no standardization of data as the QC (person who enters data) was often inexperienced, some being clinical students. It was decided against stripping data from the databases due to the Governance issues as well as the paramount amount of money that needed to be spent on it to complete the data.

The solution used was the clinical trials approach to train researchers to recruit patients from the various databases to complete clinical report forms, using a standard set of guidelines[14]. A help desk facility was also established to ensure that everyone extracting medical records knew what they were doing. Once the information was collected, information was sent to a manager and every form was monitored for irregularities. Any query with the data was sent to the center to ensure that data being entered into the database is complete. It was found that this method produced good quality data; however this approach was only possible as the number of patients sampled was small since the data collected was from a single disease and was collected from a single state. This method was therefore not recommended for a state-wide implementation.

A more efficient method needs to be found to replicate what was achieved in the AOCS on a state-wide basis, whilst maintaining data quality. One of the major problems found in integrating biobanks is the variance in the data dictionaries, created as a result of multiple cancer types. From this study, it was found that data definitions were especially important as it was difficult to cross check data types (Figure 3).

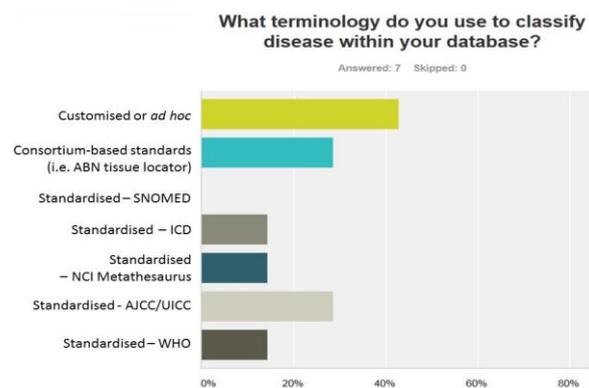

*Figure 3: Summary of terminologies used*

*B. Eighth Speech: The Victorian Experience*

Dr. Carmel Murone presented "The Victorian Experience". She is a research scientist who has worked in the Tumour Targeting program of the Ludwig Institute for Cancer Research since 1998 as well as maintaining the Austin Health Tissue Bank and fostering its transition to become a member of the Victorian Cancer Biobank.

Dr. Murone discussed how the four main independent biobanks in Victoria were integrated. All biobanks used a standard patient consent form, as well as standard operating procedures in collecting tissues. One of the biggest incentives for all sites to agree to integration was government funding. The backing of the Victorian government was critical in the success of the project, which the government did to promote local, national and international collaboration. The integration of the 4 biobanks reduced duplication of resources for collection and administration of specimen samples.

The operation model was based on a "hub and spokes model". The four independent tissue banks jointly designed the standard operating procedures, and four biospecimen banks were used to store the biospecimen samples from the 27 hospitals around them. Researchers can now lodge a single application to the Cancer Council Victoria Group for all biospecimens registered with the group in comparison to the previous system where each application had to be made to each individual hospital who owned the biospecimen, which then needed to be approved by the respective hospital ethics board. The inventory now has over 400,000 biospecimens from over 20,000 donors, and supports 140 research projects by providing more than 28,000 samples to researchers locally, nationally and internationally. It was also found that researchers are requesting more clinical management and outcome data to fully interpret their research findings, due to the improvement in infrastructure in Victoria.

Future improvements to the database include linking digital images with biospecimen data, making it web-based, holding information about the application lodged, and an invoicing model for cost recovery. NSW needs to adopt a similar stance in unifying our biobanks under one body to simplify the process of obtaining biospecimen samples to improve healthcare.

*C. Ninth Speech: What OHMR wants*

Dr. Tony Penna is the Director at the Office for Health and Medical Research (OHMR) and presented "What OHMR wants". Dr. Penna discussed OHMR's recent research in NSW biobanks and how OHMR has worked with research stakeholders including LHDs, CINSW, MRIs, and other Universities in addressing and implementing recommendations to ensure the sustainability for existing research assets with a particular focus on biobanking, bioinformatics, population-based cohort studies and record linkage. OHMR is also working closely with the cancer institute NSW to develop a state-wide biobanking framework, with the aim to improve the accessibility to samples and data.

*D. Tenth Speech: IT Options and Issues*

Jeff Christensen from Intersect discussed the IT options and issues with biobanking. Intersect's role was to provide a state-wide approach to high level IT work to provide a window to national programs. They have researched six biobanks and found that the operational procedures of each of the biobanks and IT platforms are fairly similar. The question is scope and how hospitals can be linked from different jurisdictions. It was found that databases sharing data would be a more realistic integration approach as opposed to all sites using one central system. Several issues found when integrating biobank IT systems are firewalls, network bandwidth/traffic charges (especially for hospitals when uploading large images).

*E. Panel Discussion*

The meeting concluded with a panel discussion which provided the opportunity for attendees to question the panel on issues with biobanking in NSW. Answers were provided by not only the panel but other members of the audience.

One topic of discussion was the barriers stopping Australia from the registration and accreditation of biobanks on a national level[15]. The resolution was that there was not a trusted body recognized nationally that could certify or police the system. A suggestion was for Australia to hold a national workshop to bring together biobank researchers and stakeholders to establish a trusted central body.

Another issue discussed was the need for web-based systems for biobanks so that researchers can request data online through a standardized form. Westmead has already implemented this web-based system where researchers lodge an expression of interest form which is filled out online. Emails are then sent automatically to administrative/access committee for approvals. It would be desirable for a state-wide implementation of this web-based system in NSW.

Another question was how NSW could promote/make visible the samples that are already stored in our biobanks and make it readily available to researchers. Some solutions NSW could follow include MIH which has a biospecimen resource locator, or the Canadian model which has a public registry and which biobanks can voluntarily join which is part of their certification program[16]. Another reason for the reduced visibility of samples is in the search functions for biospecimen databases. Care must be taken to ensure a simpler search function is used as opposed to a complex function which tends to freeze the system or produce no results. Currently the best option when obtaining biospecimen samples is for researchers to communicate with the biobank directly as it provides the opportunity for biobank staff to question researchers and determine the most appropriate biospecimen sample.

## IV. CONCLUDING REMARKS

In summary, the workshop was very successful. We had a number of speeches from different perspectives on how biobanking practices and informatics could be improved in NSW and Australia as a whole. We also had two very informative panel discussions which provided the opportunity for all who attended to ask questions and receive responses from not only the panel but other members of the audience. The presentations of the workshop can be downloaded from the following link - http://tcrn.unsw.edu.au/tcrn-workshop-biobanking-informatics-nsw-2013-wbin13.